\documentclass[a4paper]{jpconf}
\usepackage{graphicx,amsmath,amssymb}

\newcommand{\AIP}{{\it AIP Conf. Proc.} }
\newcommand{\CP}{{\it Chin. Phys.} }
\newcommand{\EPJ}{{\it Eur. Phys. J.} }
\newcommand{\IJMP}{{\it Int. J. Mod. Phys.} }
\newcommand{\JHEP}{{\it JHEP} }
\newcommand{\JPCS}{{\it J. Phys.: Conf. Ser.} }
\newcommand{\PoS}{{\it PoS} }
\newcommand{\PRPT}{{\it Phys. Rep.} }
\newcommand{\PREP}{{\it Preprint\/} }

\begin{document}

%%%%%%%%%%%%%%%%%%%%%%%%%%%%%%%%
%	Title
%%%%%%%%%%%%%%%%%%%%%%%%%%%%%%%%
\title{The Joint Physics Analysis Center: Recent results}

%%%%%%%%%%%%%%%%%%%%%%%%%%%%%%%%
%	Authors
%%%%%%%%%%%%%%%%%%%%%%%%%%%%%%%%
\author{C\'esar Fern\'andez-Ram\'{\i}rez}
\address{Instituto de Ciencias Nucleares, 
Universidad Nacional Aut\'onoma de M\'exico,
Ciudad de M\'exico 04510, Mexico}
\ead{cesar.fernandez@nucleares.unam.mx}

%%%%%%%%%%%%%%%%%%%%%%%%%%%%%%%%
%	Abstract
%%%%%%%%%%%%%%%%%%%%%%%%%%%%%%%%
\begin{abstract}
We review some of the recent achievements of the Joint Physics Analysis Center, 
a theoretical collaboration with ties to experimental collaborations, that aims
to provide amplitudes suitable for the analysis of the current 
and forthcoming experimental data on hadron physics. 
Since its foundation in 2013, the group is focused on hadron spectroscopy in preparation for the 
forthcoming high statistics and high precision experimental data from
BELLEII, BESIII, CLAS12, COMPASS, GlueX, LHCb and (hopefully) PANDA collaborations.
So far, we have developed amplitudes for 
$\pi N$ scattering, $\bar{K}N$ scattering,  
pion and $J/\psi$ photoproduction, two kaon photoproduction
and three-body decays of light mesons ($\eta$, $\omega$, $\phi$). 
The codes for the amplitudes are available to download from the group web page and
can be straightforwardly incorporated to the analysis of the experimental data.
\end{abstract}

%%%%%%%%%%%%%%%%%%%%%%%%%%%%%%%%
%	Introduction
%%%%%%%%%%%%%%%%%%%%%%%%%%%%%%%%
\section{Introduction}
One of the many challenges of physics is to achieve a comprehensive understanding
of the phenomena produced by strong coupling Quantum Chromodynamics (QCD), 
responsible for the binding of quarks inside hadrons and 
for more than 95\% of the mass of the visible Universe.
Hadrons are color singlets due to color confinement. 
Hence, the simplest structures are $q\bar{q}$ (quark-antiquark) for mesons
and $qqq$ (3-quarks) for baryons. However, there are many other ways 
we could build color neutral hadrons, for example:
glueballs made out of a number of gluons, multi-quark states such as tetraquarks ($qq\bar{q}\bar{q}$) 
or pentaquarks ($qqqq\bar{q}$), hybrid mesons and baryons with gluonic 
components ($q\bar{q}g$ and $qqqg$) and so on. For a recent, brief and pedagogical 
summary of the current status of hadron physics we refer the reader to \cite{MRP2016a}.

In the last years we have witnessed a dramatic advancement 
in detection techniques and accelerator technologies
from the experimental side and algorithms for first principle QCD analyses 
on the theoretical side \cite{ATHOS}.
Consequently, many experimental facilities throughout 
the world have research programs on hadron spectroscopy, among them: 
LHCb \cite{LHCb}, CMS \cite{CMS} and COMPASS \cite{COMPASS} at CERN (Geneve, Switzerland),
CLAS12 \cite{CLAS} and GlueX \cite{GlueX} at Jefferson Lab (Newport News VA, USA),
BESIII  \cite{BEPC} at the Beijing Electron Positron Collider (Beijing, China),
KLOE2 \cite{KLOE2} at Laboratori Nazionali di Frascati (Frascati, Italy),
BELLEII \cite{BELLE2} at KEK (Tsukuba, Japan),
the Hadron Experimental Facility  \cite{JPARC}  at J-PARC (Tokai, Japan), 
and (hopefully) the future PANDA \cite{PANDA} at FAIR (Darmstadt, Germany).
As a result, several candidates for {\it exotic} hadrons beyond the $q\bar{q}$ and $qqq$ 
picture have been discovered both in the meson \cite{tetraquark}
and baryon \cite{LHCbpentaquark} sectors.
 
Given the experimental and theoretical interest on hadron spectroscopy, the
Joint Physics Analysis Center (JPAC) \cite{JPACWebpageIndiana} was set up to 
develop theoretical and phenomenological analysis methods to support hadron physics experiments.
It started out in 2013 with 8 researchers,
as a joint venture between Indiana University, George Washington University and Jefferson Lab.
Currently it has expanded to 20 researchers distributed among 
the three founding institutions plus
Rheinischen Friedrich-Wilhelms Universit\"at Bonn
and Johannes Gutenberg Universit\"at Mainz in Germany, 
IFIC/CSIC-Universidad de Valencia in Spain, 
and Universidad Nacional Aut\'onoma de M\'exico in Mexico.

The JPAC researchers are using S-matrix theory principles, 
--{\it i.e.} analyticity, crossing symmetry and unitarity \cite{Eden}-- 
to develop scattering amplitudes for 
several hadron reactions of theoretical and experimental interest
In doing so, we work closely with experimentalists to implement the amplitudes 
in the existing data-analysis software.

The methods to write the amplitudes include the K-matrix \cite{kmatrix} and
N/D parametrizations \cite{noverd}. These can be complemented with chiral symmetry constraints
at low-energies \cite{chiral} or Regge asymptotics  \cite{PDBCollins} if needed.
The amplitudes can be analytically continued from the real axis where the experimental data {\it live}
to the complex-$s$ plane and to the unphysical Riemann sheets, which allows us to unravel
the existing resonances contributing to the physical reactions.
It is important to note that the amplitudes developed 
by JPAC and the analytical continuation to the complex plane 
are also necessary to extract the resonances and their  properties
from first principle lattice gauge QCD \cite{lattice}. 

In the next section we briefly summarize some of the results obtained 
by the group during the last two years. 

%%%%%%%%%%%%%%%%%%%%%%%%%%%%%%%%
%	Recent Results
%%%%%%%%%%%%%%%%%%%%%%%%%%%%%%%%
\section{Recent results}

\subsection{Three-body meson decays}
Three-body decays play an essential role studying hadron dynamics and hadron spectroscopy.
The extraction of CP violation phases often requires the amplitude analysis of three-body $B$ and $D$ decays
\cite{BDdecays}
and several candidates for non-quark model states have been observed in 
heavy mesons decaying into three particles \cite{LHCbpentaquark,heavymesondecay,PDG2014}. 
Heavy-meson decays have complicated phase spaces and their analysis can be cumbersome,
hence, it is better to test our ideas on amplitude-building with simpler cases, 
like the decay of light mesons, {\it i.e.} $\eta$, $\omega$, $\phi$ $\to 3\pi$.
Besides, these decays are interesting {\it per se} given the current and forthcoming precision data 
from various collaborations \cite{lightmesondecay}.
For these reasons, at JPAC we have developed an intensive work on
dispersive approaches for three-particle final state interaction based on 
the Khuri-Treiman equations \cite{PG2015a}
and generalized Veneziano amplitudes \cite{APS2014a}, 
---this last work for $J/\psi \to 3\pi$ decay, although it can also be used for light-meson decays.
Based on the dispersive approach we have developed amplitudes for $\eta$ \cite{PG2015b,PG2016a}, 
$\omega$ \cite{IVD2015a}, and $\phi$  \cite{IVD2015a} decays to $3\pi$.

The $\eta \to 3\pi$ decay is of particular interest because it is isospin-breaking
and provides insight on the light-quark mass difference
\begin{equation}\label{Eq:Q}
\frac{1}{Q^2}=\frac{m_d^2-m_u^2}{m_s^2-\hat m^2}\,,\quad \hat m =\frac{(m_u+m_d)}{2}\,,
\end{equation}
where $m_u$, $m_d$ and $m_s$ are the masses of $u$, $d$ and $s$ quarks respectively.
In \cite{PG2016a} 
we performed a simultaneous global fit to the  KLOE-2 \cite{Adlarson:2014aks} 
and WASA-at-COSY data \cite{Anastasi:2016qvh} with two real parameters and we
determined $Q = 21.6 \pm 0.4$  \cite{PG2016a}
by matching our dispersive approach to nex-to-leading 
order Chiral Perturbation Theory ($\chi PT$) near the Adler zero \cite{EtaPassemar}.

\subsection{Meson-baryon scattering}
Meson-baryon scattering is one of the main resources to study the baryon spectrum. 
At JPAC, we have focused on $\pi N$ \cite{VM2015b} and $\bar{K}N$ \cite{CFR2016a} scattering
because these are common final states detected in hadron experiments.
In \cite{VM2015b} we developed high-energy $\pi N$ amplitudes using Regge physics \cite{PDBCollins} 
and connected them to the low-energy amplitudes (resonance region) from \cite{RLW2012} 
using dispersion relations and Finite-Energy Sum Rules (FESRs).
Consequently, we built a new set of $\pi N$ amplitudes that cover the whole energy range for forward scattering.

In \cite{CFR2016a} we built a coupled-channel $\bar{K}N$ scattering model in the resonance region
that incorporates up to 13 channels per partial wave and fulfills 
unitarity,  has the correct analytical properties for the amplitudes in the resonance region
and has the right threshold behavior for the partial waves. 
We also imposed analyticity of partial waves 
in the complex angular momentum plane \cite{Gribov}.
The model was fitted to the Kent State University single-energy partial waves \cite{Manley13a}
using a genetic algorithm \cite{genetic} and \textsc{MINUIT} \cite{MINUIT}. 
We estimated the uncertainties in the parameters using bootstrap technique \cite{NumericalRecipes}.
Once the model was fitted to the single-energy partial waves and the parameters 
established, we analytically continued
the partial waves to the unphysical Riemann sheets and found the $\Lambda^*$ and $\Sigma^*$ poles (resonances)
that constitute the low-lying hyperon spectrum \cite{CFR2016a,CFR2016d}.
We expect to develop forward-scattering $\bar{K}N$ amplitudes  \cite{VM2016b} 
that cover the whole energy range
combining high-energy $\bar{K}N$ Regge amplitudes with the low-energy amplitudes in \cite{CFR2016a}
using FESR as was done in \cite{VM2015b} for $\pi N$ scattering.

The identification of baryons beyond the constituent quark model 
(either hybrids with glue as an essential constituent, molecules or pentaquarks) 
requires the identification of a whole flavor family, and not only 
the detailed study of one particular state \cite{MRP2016a}.
Resonances of different angular momenta share the same Regge trajectory, 
what gives the spectrum an organized structure \cite{PDBCollins,CFplot}.
The structure of the spectrum and the shape of the trajectories provide information on the composition of the states
\cite{CFR2016a,CFR2016d,mesonregge,CFR2016b,CFR2016c}.
In  \cite{CFR2016b} we combined Regge phenomenology \cite{PDBCollins,Gribov},
the information on the hyperon spectrum from \cite{CFR2016a},
and the $\Lambda$(1405) pole positions from \cite{Lambda1405poles} to
gain insight on the nature of the two $\Lambda$(1405) states.
We found that the higher-mass pole appears to be (mostly) a three-quark hyperon and the lower-mass pole
either a molecule or a pentaquark \cite{CFR2016b,CFR2016c}.

\subsection{Meson photoproduction}
Another front where JPAC is producing results is meson photoproduction. 
In particular, we have worked on two kaon photoproduction \cite{MS2015a}, 
high-energy pion photoproduction \cite{VM2015a}
and $J/\psi$ photoproduction  \cite{ANHB2016a}.

In \cite{MS2015a} the double-Regge limit for the $\gamma p \to K^+ K^- p$ reaction was computed.
This reaction has been measured by CLAS collaboration (currently under analysis) 
and is of interest to study mesons with hidden strangeness
that decay into two kaons. The double-Regge production mechanism constitutes a background to the reaction
that contributes to the whole Dalitz plot and needs to be removed in order to isolate the mesons of interest.

There are many pion photoproduction models in the literature that describe 
the resonance region \cite{photoproduction}
but not recent ones for the high-energy regime.
Considering that CLAS and GlueX collaborations will collect 
pion photoproduction data in the $E_\gamma \simeq 3-10$ GeV region 
we developed a high-energy neutral pion photoproduction \cite{VM2015a}
that covers the $E_\gamma > 4$ GeV region where the Regge poles dominate the amplitudes. 

LHCb collaboration reported two possible candidates for pentaquarks in the 
$J/\psi p$ channel \cite{LHCbpentaquark},
however, the existence of these states is still to be confirmed. One of the most interesting proposals to detect the 
narrow $P_c(4450)$ pentaquark was made by Karliner and Rosner in \cite{MK2016a}.
They proposed to measure $J/\psi$ photoproduction from the proton.
If there is a pentaquark resonance that strongly couples to the  $J/\psi p$ channel, 
it should be possible to detect it in the $\gamma p \to J/\psi p$ reaction.  
In \cite{ANHB2016a} we built up a full model for this reaction and analyzed the available experimental data on 
$J/\psi p$ photoproduction from the proton. The resonance was introduced as a Breit--Wigner 
amplitude and the background was modeled with a Pomeron exchange. 
We found that data allow for the existence of a pentaquark whose $J/\psi p$ 
branching ratio has to be below 30\% for spin-parity assignment $J^P=3/2^-$ 
and below 17\% for $J^P=5/2^+$ at a 95\% confidence level. 
Jefferson Lab has just approved an experiment to measure this reaction \cite{PcJLab}.

%%%%%%%%%%%%%%%%%%%%%%%%%%%%%%%%
%	Web page
%%%%%%%%%%%%%%%%%%%%%%%%%%%%%%%%
\section{Web page}
We created an interactive website to make the exchange of information between 
JPAC and other theorists and experimentalist easier \cite{VM2016a}.
The JPAC web page at Indiana University \cite{JPACWebpageIndiana} hosts 
downloadable versions of the codes to compute
the $\eta \to 3 \pi$ amplitude from \cite{PG2015b,PG2016a},
the $\omega/\phi \to 3 \pi$ amplitudes from \cite{IVD2015a},
the $\pi N$ amplitudes employing FESRs from \cite{VM2015b},
the $\bar{K}N$ partial waves and observables in the resonance region from \cite{CFR2016a}
and the high-energy pion photoproduction observables from \cite{VM2015a}.
The codes can also be run on the web page and the outputs downloaded.
Codes to compute the generalized Veneziano amplitude from \cite{APS2014a} 
and the double-Regge limit for the $\gamma p \to K^+ K^- p$ reaction using the model in \cite{MS2015a} 
are available upon request.
The codes for pentaquark photoproduction  \cite{ANHB2016a} will be available in the near future.
We are currently in the process of building up a mirror of the JPAC web page with additional functionalities
at Universidad Nacional Aut\'onoma de M\'exico,
hosted at the web servers of  Instituto de Ciencias Nucleares \cite{JPACWebpageUNAM}.

%%%%%%%%%%%%%%%%%%%%%%%%%%%%%%%%
%	Final remarks
%%%%%%%%%%%%%%%%%%%%%%%%%%%%%%%%
\section{Final remarks}
The JPAC is a theoretical collaboration oriented to work with experimental collaborations 
and willing to cooperate with other theory groups. 
At present we have standing collaborations with BESIII, COMPASS, CLAS, GlueX, KLOE and LHCb.
We are focused on hadron spectroscopy delivering amplitudes for 
three-body meson decays \cite{PG2015a,APS2014a,PG2015b,PG2016a,IVD2015a},
meson-baryon scattering \cite{VM2015b,CFR2016a,CFR2016d,VM2016b},
Regge phenomenology to gain insight on the internal structure of hadrons \cite{CFR2016b,CFR2016c}
and meson photoproduction \cite{MS2015a,VM2015a,ANHB2016a}.
Codes for the amplitudes are available either from the  
JPAC web page \cite{JPACWebpageIndiana,VM2016a,JPACWebpageUNAM}
or upon request. We invite the hadron physics community 
to visit the website and send their comments and suggestions to the JPAC members.

%%%%%%%%%%%%%%%%%%%%%%%%%%%%%%%%
%	Ack
%%%%%%%%%%%%%%%%%%%%%%%%%%%%%%%%
\ack
We thank the Division of Particles and Fields of the Mexican Physical Society for 
their kind invitation to write this brief report on JPAC purpose and recent activities.

%%%%%%%%%%%%%%%%%%%%%%%%%%%%%%%%
%	References
%%%%%%%%%%%%%%%%%%%%%%%%%%%%%%%%

\end{document}